 \documentclass[preprint,review,12pt]{elsarticle}

\usepackage{graphicx}
\usepackage{amssymb}
\usepackage{bm}
\usepackage{multirow}

\journal{Journal of Non-Crystalline Solids}

\begin{document}

\begin{frontmatter}


\title{Morphology and magnetic properties of nanocomposite magnetic multilayers {[(Co$_{40}$Fe$_{40}$B$_{20}$)$_{34}$(SiO$_2$)$_{66}$]/[C]}$_{47}$}

\ead{email address}

\author[label1,label2]{V.\,Ukleev}
\ead{ukleev@lns.pnpi.spb.ru}
\address[label1]{National Research Centre "Kurchatov Institute" B. P. Konstantinov Petersburg Nuclear Physics Institute, 188300 Gatchina, Russia}
\address[label2]{Saint-Petersburg Academic University — Nanotechnology Research and Education Centre of the Russian Academy of Sciences, 194021 Saint-Petersburg, Russia}

\author[label1]{E. Dyadkina}

\author[label3]{A. Vorobiev}
\address[label3]{Department of Physics and Astronomy, Uppsala University, Box 516, SE-75120 Uppsala, Sweden}

\author[label1]{O. V. Gerashchenko}

\author[label4,label5]{L. Caron}
\address[label4]{Delft University of Technology, 2628 CN Delft, Netherlands}
\address[label5]{Max Planck Institute for Chemical Physics of Solids,
Nöthnitzer Straße 40, D-01187 Dresden, Germany}

\author[label6]{A. V. Sitnikov}

\author[label6]{Yu. E. Kalinin}
\address[label6]{Voronezh State Technical University, 394026 Voronezh, Russia}

\author[label1,label7]{S. V. Grigoriev}
\address[label7]{Saint-Petersburg State University, 198904 Saint-Petersburg, Russia}

\begin{abstract}
We report on the investigation of morphology, magnetic and conductive properties of the mutilayered nanostructures [(Co$_{40}$Fe$_{40}$B$_{20}$)$_{34}$(SiO$_2$)$_{66}$]/[C]$_{47}$ consisting of the contacting magnetic (Co$_{40}$Fe$_{40}$B$_{20}$)$_{34}$(SiO$_2$)$_{66}$ nanocomposite and amorphous semiconductor carbon C layers.  It is shown by Grazing-Incidence Small-Angle X-ray Scattering method that the ordering and the size of nanoparticles in the magnetic layers do not change profoundly with increasing of carbon layer thickness. Meanwhile, the electrical conductance and the magnetic properties are significantly varied: resistance of the samples changes by four orders of magnitude and superparamagnetic blocking temperature changes from 15 K to 7 K with the increment of carbon layer thickness $h_c$ from 0.4 nm to 1.8 nm. We assume that the formation of the homogeneous semiconductor interlayer leads to modification of the metal-insulator growth process that drives the changes in the magnetic and conductive properties.
\end{abstract}


\end{frontmatter}

\section{Introduction}
\label{S:1}

Metal-insulator nanocomposites, which are metallic granules embedded into amorphous insulating matrix have attracted large interest due to the their soft magnetic performance in high-frequency electromagnetic regions \cite{Ikeda2002, hfr, 951146}, magnetotransport phenomena, such as magnetoresistance and giant Hall effect \cite{Pakhomov1995, Zhao1997, Vovk2002, Wan2002, Liu2004}, as well as their structural stability \cite{Cho2002,Chun2000,Choi2005} because of an absence of grain boundaries and smooth interfaces. Further, due to the significant development of micro-, nanoelectronics and spintronics, many studies are currently devoted to the multilayered nanostructures containing the contacting ferromagnetic and semiconductor (SC) layers \cite{Wolf2001,Fert2001}.

In multilayered systems the morphology of one layer in the periodic stack is linked with the neighboring layers and affects on the entire properties of the system. The morphology of metal-insulator (MI) layers is determined by their thickness and the metal concentration. At the same time, SC layers are chemically homogeneous, and their morphology is determined only by their thickness and roughness. Recently it was shown that the conductive and magnetic properties of MI / SC nanostructures are determined both by the composition and morphology of MI \cite{Iskhakov2007} and SC layers \cite{Dyadkina2011, Dunets2013, Dyadkina2014}. Some of the recent works were aimed to study an indirect coupling between the nanoparticles in MI layers through semiconducting interlayers. It was shown that such coupling takes place in [(Co$_{40}$Fe$_{40}$B$_{20}$)$_{50}$(SiO$_2$)$_{50}$]/[$\alpha$-Si]$_{60}$ and exchange field was estimated: $A_s \approx 0.15 \times 10^{-6}$ erg/cm \cite{Komogortsev2013}. On the other hand the polarized neutron reflectometry in (Co$_{45}$Fe$_{45}$Zr$_{20}$)$_{35}$(Al$_2$O$_3$)$_{65}$]/[$\alpha$-Si]$_{36}$ demonstrated the magnetic degradation of the interface MI / SC \cite{Dyadkina2011,Dyadkina2014}. Therefore the mechanism and characteristics of the indirect magnetic interaction between nanoparticles through the SC is still uncertain.

One of the perspective objects for the investigation of the magnetic properties is the multilayered system {[(Co$_{40}$Fe$_{40}$B$_{20}$)$_{34}$(SiO$_2$)$_{66}$]/[C]}$_{47}$ where the number of bilayers is equal to 47. Each bilayer of the nanostructure consists of a layer of an amorphous MI composite (Co$_{40}$Fe$_{40}$B$_{20}$)$_{34}$(SiO$_2$)$_{66}$ and an adjacent SC layer is made of amorphous carbon C. Amorphization of these materials was confirmed by X-ray diffraction. The amorphous alloy Co$_{40}$Fe$_{40}$B$_{20}$ is used as the magnetic component of the composite (Co$_{40}$Fe$_{40}$B$_{20}$)$_{34}$(SiO$_2$)$_{66}$ because it can be easily amorphized and contains 80 at. \% of the ferromagnetic phase. The percolation threshold between metallic and insulator phases of the nanocomposite (Co$_{40}$Fe$_{40}$B$_{20}$)$_{x}$(SiO$_2$)$_{(1-x)}$ was found for $x_c = 30$ at. \% \cite{Iskhakov2007}. The behavior of in-plane electrical resistivity $\rho$ of the multilayered nanostructure in dependence on carbon layer thickness at different temperatures was studied in Ref. \cite{Dunets2013}. It was shown, that resistivity of the nanostructre decreases by four orders of magnitude with carbon layer thickness $h_c$ growth from 0.4 nm to 1.8 nm (Fig.\ref{Fig1}a). 

The objective of our present work is to study how the morphology of the carbon layer thickness affects the structural features and the related magnetic properties of the nanostructure [(Co$_{40}$Fe$_{40}$B$_{20}$)$_{34}$(SiO$_2$)$_{66}$]/[C]$_{47}$ by the  combination of experimental methods: Grazing-Incidence Small-Angle X-ray Scattering (GISAXS) and Superconducting Quantum Interference Device (SQUID) magnetometry.

\section{Samples}
\label{S:2}

The nanostructures were fabricated by ion-beam co-sputtering of two targets on a rotating glass-ceramic substrate. One target is the vacuumcasted alloy Co$_{40}$Fe$_{40}$B$_{20}$ with silicon oxide SiO$_2$ plates fixed on its surface. Another target is amorphous carbon C. The composition of MI was monitored by the X-ray fluorescence analysis. All peculiarities of the sputtering method, choice of components and control of the sample parameters are described in Refs. \cite{Zolotukhin2002, Iskhakov2007, Stognei2010, Dunets2013}. 

In the present study we investigated two samples with the concentration of metal $x=34.0 \pm 1.5$ at. $\%$ (i.e. just above the percolation threshold) and with different carbon layer thicknesses $h_c$ known from the deposition time. The samples with $h_c=0.40 \pm 0.18$ nm and $h_c=1.80 \pm 0.12 $ nm are notified $S_1$ and $S_2$, respectively. 

\section{Experiment}
\label{S:3}

\begin{figure}
\includegraphics[width=8.5cm]{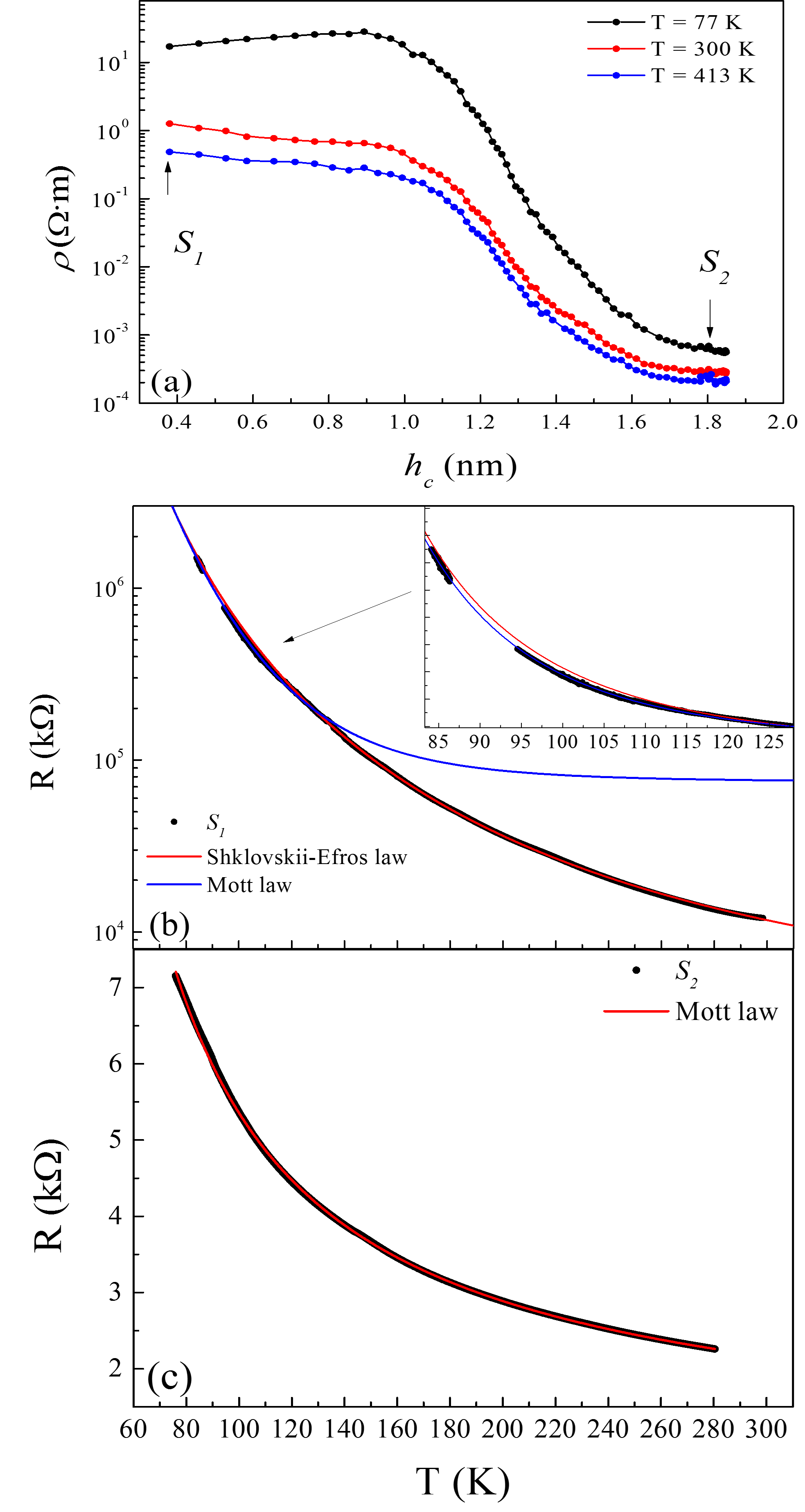}
\caption{(a) In-plane electrical resistivity $\rho$ of the multilayer nanostructures [(Co$_{40}$Fe$_{40}$B$_{20}$)$_{34}$(SiO$_2$)$_{66}$]/[C]$_{47}$ vs. carbon layer thickness $h_c$ at different temperatures $T = 77$, 300 and 413 K. (b) The temperature dependence of the measured electrical resistance (symbols) and simulated (lines) using the exponential laws (lines) for the sample $S_1$. The inset shows the crossover region between the Shklovskii-Efros and Mott laws. (c) Measured (symbols) and simulated (line) resistance vs. temperature for the sample $S_2$.}
\label{Fig1}
\end{figure}

The standard four-probe measurement with the correction on the sample size ($Y_1 = 11$ mm for the sample $S_1$, $Y_2 = 5$ mm for the sample $S_2$) and distance between the electrodes ($W_1 = 5$ mm for the sample $S_1$, $W_2 = 3$ mm for the sample $S_2$) was used to probe the transport properties of the multilayered samples. Measurements of the resistivity have shown four orders of difference for the samples $S_1$ and $S_2$ (Fig. \ref{Fig1}). The temperature dependence of the in-plane resistance for both samples is well described by the exponential law for the hopping conductivity \cite{mott}:
\begin{equation}
R = R_0 e^{(\frac{T_0}{T})^{\gamma}},
\label{rt}
\end{equation}
where $R_0$ and $T_0$ are constants. The parameter $\gamma$ is defined by the mechanism of conductivity: in case of charge tunneling transport $\gamma = 1/2$ (Shklovskii-Efros law \cite{efros1975coulomb}) and in case of strong Anderson localization of the electronic states at Fermi level due to disorder $\gamma = 1/4$ (Mott law \cite{mott}). The experimental results are shown in Fig.\ref{Fig1}b,c. The temperature behavior of high-Ohmic sample $S_1$ can be satisfactorily described by the Shklovskii-Efros law $R\sim exp(T_0/T)^{1/2}$ at low temperature. Crossover between Shklovskii-Efros and Mott conductivity $R\sim exp(T_0/T)^{1/4}$ regimes can be observed at the temperature $T \approx 135$ K (see inset in Fig. \ref{Fig1}b). On the other hand, the resistance of the low-Ohmic sample $S_2$ can be uniformly described by the single Mott law exponent and no crossover takes place (Fig. \ref{Fig1}c). 

\begin{figure}
\includegraphics[width=8cm]{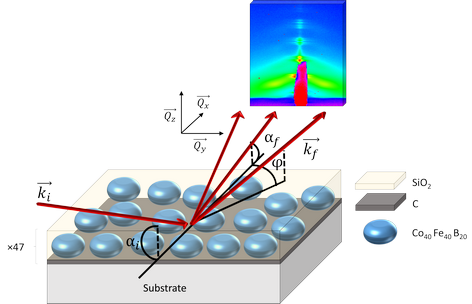}
\caption{The geometry of GISAXS experiment: \textbf{$k_i$} and \textbf{$k_f$} are the wavevectors of the incident and scattered beams, respectively.}
\label{Fig2}
\end{figure}

To understand the connection between the multilayered structure morphology and large variation of the electrical properties in dependence on the SC layer thickness Grazing-Incident Small-Angle X-ray Scattering was used. GISAXS allows one to analyze the distribution of the inhomogeneities (in our case, metal nanoparticles), their size and spatial organization. Grazing-incident geometry improves the surface sensitivity compared to the classical SAXS transmission geometry making this technique suitable for investigation of surfaces, interfaces, films, and multilayered nanostructures. By changing the angle of incidence ($\alpha_i$) near the critical angle ($\alpha_c$) of the total external reflection, it is possible to regulate the spread of the beam into the sample and to obtain the in-depth distribution of the inhomogeneties, as well as their size and shape. The  geometry of GISAXS experiment is determined by grazing-incidence angle $\alpha_i$ of the synchrotron radiation beam on the sample surface and two scattering angles, $\alpha_f$ and $\varphi$ (Fig. \ref{Fig2}). The angles $\alpha_i$, $\alpha_f$ and $\varphi$ determine the values of components of the momentum transfer vector: $Q_x$, $Q_y$, $Q_z$.

Thus, by measuring the component of the momentum transfer $Q_z$ perpendicular to the sample plane, one can get the information on the distribution of electron density in $z$ direction, while by measuring  the component $ Q_{||}$ in the sample plane $(x, y)$, it is possible to study the lateral structure of the sample. The components of the momentum transfer can be expressed through the incidence  and scattering angles \cite{Renaud2009}:

\begin{equation} 
Q_{\rm z}(\varphi , \alpha_{\rm f}) =  \frac{2\pi}{\lambda} \left( \sin \alpha_{\rm f}  + \sin \alpha_{\rm i} \right),
\end{equation}
$$Q_{||}(\varphi , \alpha_{\rm f}) =  \sqrt{Q_{\rm x}^2(\varphi , \alpha_{\rm f})+Q_{\rm y}^2(\varphi , \alpha_{\rm f})},$$
$$Q_{\rm x}(\varphi , \alpha_{\rm f}) =  \frac{2\pi}{\lambda} \left( \cos \alpha_{\rm f}  \cos\varphi - \cos \alpha_{\rm i} \right),$$
$$Q_{\rm y}(\varphi , \alpha_{\rm f}) =  \frac{2\pi}{\lambda} \cos \alpha_{\rm f}  \sin \varphi.$$

\begin{figure*}
\includegraphics[width=16cm]{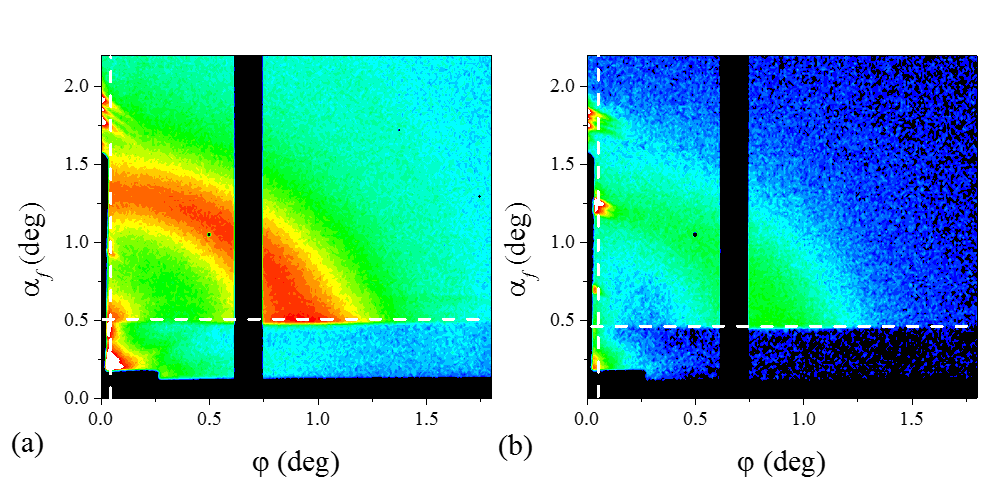}
\caption{Two-dimensional maps of GISAXS for the samples (a) $S_1$ and (b) $S_2$. The white dash-dot lines correspond to the sections presented in Fig. \ref{Fig4}, \ref{Fig5}. The shadows between $0.61^\circ < \varphi < 0.74^\circ$ correspond to dead area between the modules of Pilatus detector.}
\label{Fig3}
\end{figure*}

The GISAXS experiment was carried out at the ID10 beamline of European Synchrotron Radiation Facility (ESRF, Grenoble, France). In the experiment, the collimated $10\times10$ $\mu$m$^2$ beam of photons with the wavelength of $\lambda = 0.56$ \AA ~was used. The scattering intensity was measured by the two-dimensional position-sensitive detector Pilatus 300K with exposition time of 30 seconds for one scattering pattern. The central part of the detector was protected from the direct beam by tungsten beamstop.

Two-dimensional maps of GISAXS obtained for the samples $S_1$ and $S_2$ at $\alpha_i=0.2^\circ$ are shown in Fig.\ref{Fig3}. The shadows in Fig. \ref{Fig3} correspond to the dead area between the modules of Pilatus detector. From the distribution of the scattering intensity in the momentum transfer space \textbf {Q} ($ \alpha_i, \varphi, \alpha_f$) one can obtain the characteristic distances between nanoparticles in the real space. The period of the multilayered structure can be calculated as $ \Lambda = 2 \pi/\Delta Q_z $, where $\Delta Q_z $ is the distance between peaks along $Q_z$ at $\varphi=0^\circ$. Similarly, one can get the characteristic values of the interparticle distances $l$ in the sample plane by analyzing the scattering intensity distribution along $Q_{||}$.

\begin{figure}
\includegraphics[width=8cm]{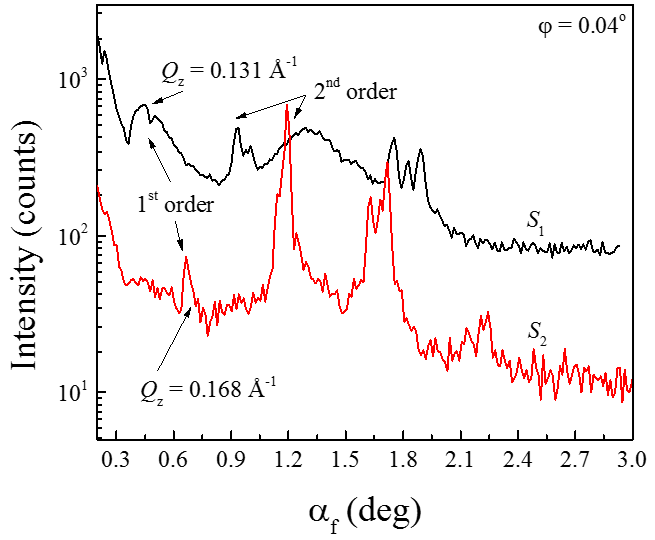}
\caption{The sections of two-dimensional intensity maps along $\alpha_{f}$ ($\varphi = const$) for the samples $S_1$, and $S_2$.}
\label{Fig4}
\end{figure}

\begin{figure}
\includegraphics[width=8cm]{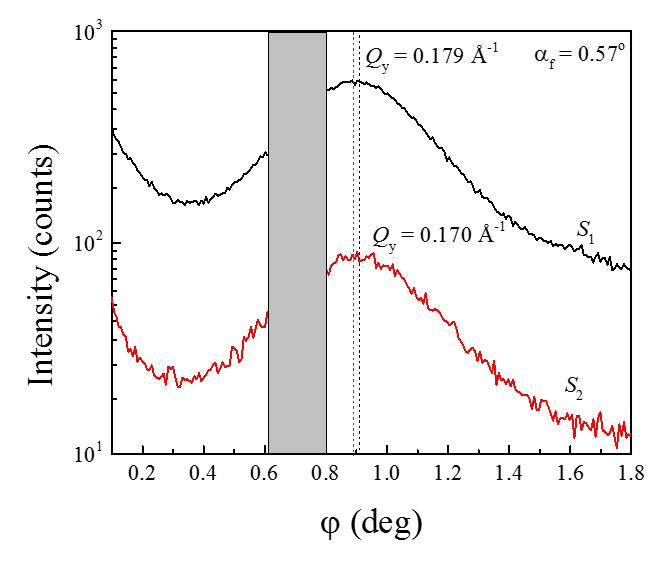}
\caption{The sections of two-dimensional intensity maps along $\varphi$ ($\alpha_{f} = const$) for the samples $S_1$, and $S_2$. The line breaking between $0.61^\circ < \varphi < 0.74^\circ$ corresponds to dead area between the modules of Pilatus detector. The dashed lines indicate the peak centers found by Gaussian function fitting.}
\label{Fig5}
\end{figure}

The generic feature of the two-dimensional maps for all the samples is at least one pronounced specular Bragg peak, that confirms the periodic stacking of MI and carbon layers. No Bragg peaks can be observed at $Q_{z}\neq 0$ values, indicating that nanoparticles are not vertically ordered \cite{Vegso2012}. The relative contribution of the diffuse background to the GISAXS signal compared to the Bragg peaks intensity is much larger in the case of the sample $S_1$ than in the case of sample $S_2$ that is a result of coherent reflection of X-rays from the well-pronounced layered structure in $S_2$. For the qualitative analysis of the layers ordering the sections of intensity along $Q_{z}$ were analyzed (Fig.\ref{Fig4}). Splitting of the Bragg peaks in higher $Q$-range is caused by the contribution of the multiple reflection from the surface roughness. To determine the  periodicity of the multilayers the most pronounced peaks were chosen. Sample $S_1$ shows the set of wide satellites which correspond to the periodic structure with $\Lambda = 4.80 \pm 0.11$ nm. For the sample $S_2$ much more pronounced Bragg peaks indicate the periodicity of the bilayers $\Lambda = 3.73 \pm 0.05$ nm. The values and error bars for $\Lambda$ was calculated by fitting Bragg peaks by the Gaussian function with additional contribution caused Fresnel decay. Thus using the nominal thickness of carbon layer known from the deposition time one can obtain the thickness of MI layers $h_{MI} = \Lambda - h_c$. It should be noted, that the intensity of the 2nd order satellite in case of the sample $S_2$ is higher than the intensity of the 1st order satellite (Fig. \ref{Fig3}). This feature indicates a contribution of the MI/SC interfaces to the scattering increasing. In our case huge off-specular scattering complicates interpretation of the data concerning this additional contribution.

The average interparticle distances in the sample plane were determined from the sections of GISAXS intensity along the $\varphi$ at the constant $\alpha_f$ (Fig. \ref{Fig5}). One can found that the thickness of carbon layer almost does not affect the peak positions, and, consequently, the lateral distribution of metallic nanoparticles in the samples plane. 

The structural parameters of the samples $\Lambda$ and $l$ obtained from the GISAXS experiments are shown in Table \ref{table}.

\begin{table*}[t]
  \centering
  \caption{The structural parameters of the samples $S_1$, $S_2$ determined by the GISAXS experiments. Data for the carbon layer thickness were taken from the deposition time.}
  \begin{tabular}{|c|p{1.8cm}|p{1.8cm}|p{1.8cm}|p{1.8cm}|p{1.8cm}|p{1.8cm}|c|p{1.8cm}|c|}
    \hline

Sample &$\Lambda$ (nm) & $h_c$ (nm) & $h_{MI}$ (nm) & $l$ (nm) & $r_h$,(nm) & $V$, (nm)$^3$ & $\Delta V$, (nm)$^3$ (SQUID)\\
    \hline
    $S_1$ & 4.80 $\pm$ 0.11 & 0.40 $\pm$ 0.18& 4.40 $\pm$ 0.29 & 3.51 $\pm$ 1.49 & 1.66 $\pm$ 0.71& 25.4 $\pm$ 15.5& 4.12 $\pm$ 0.03 \\ \hline
    $S_2$ & 3.73 $\pm$ 0.05 & 1.80 $\pm$ 0.12& 1.93 $\pm$ 0.17 & 3.69 $\pm$ 1.53 & 1.76 $\pm$ 0.73& 12.3 $\pm$ 7.4 & 5.01 $\pm$ 0.03 \\ \hline
  \end{tabular}
  \label{table}
\end{table*}

\begin{figure}
\includegraphics[width=8cm]{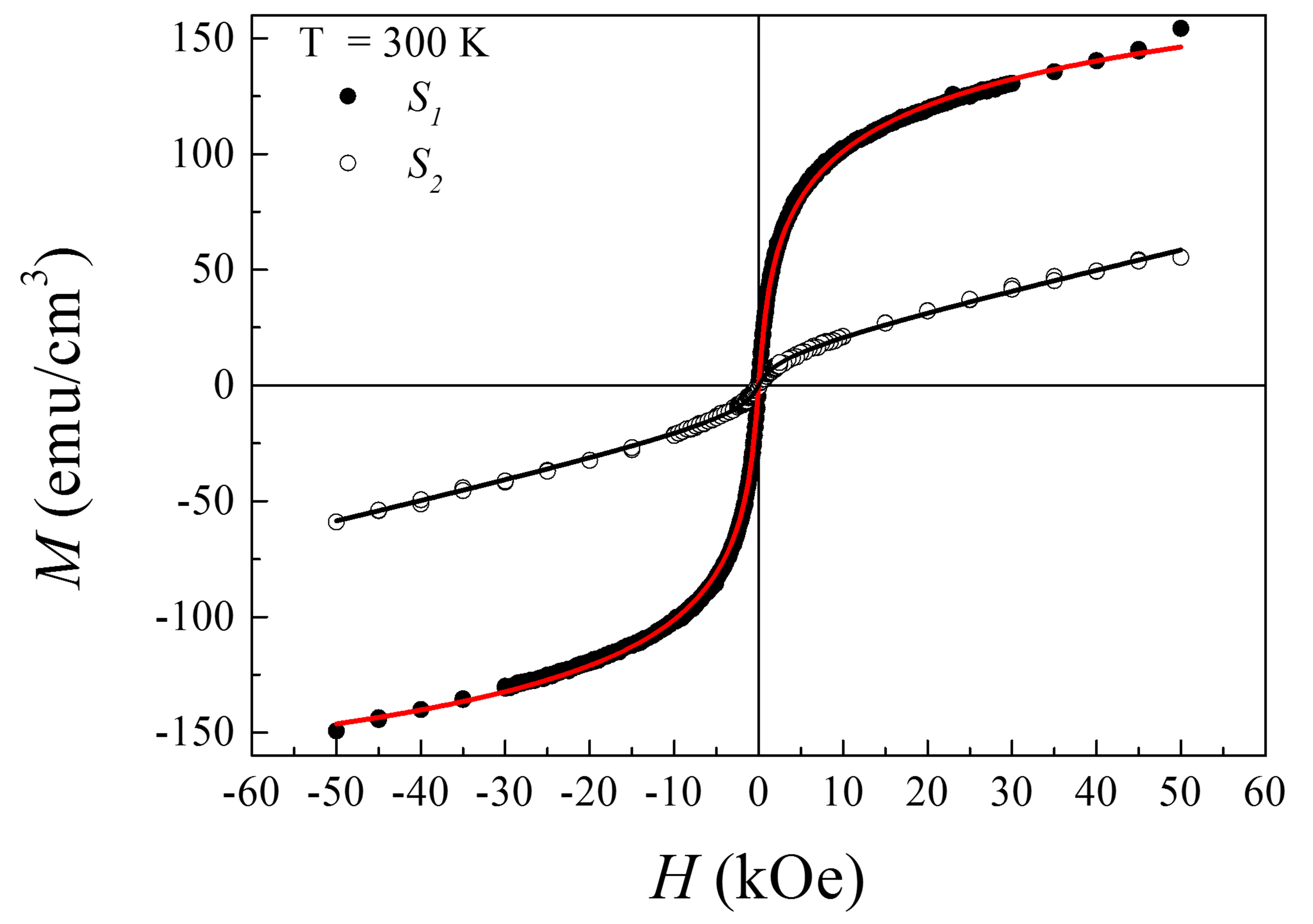}
\caption{The magnetization loops of the samples $S_1$, $S_2$ measured at room temperature. Solid lines correspond to the fitted model.}
\label{Fig6}
\end{figure}

\begin{figure}
\includegraphics[width=8cm]{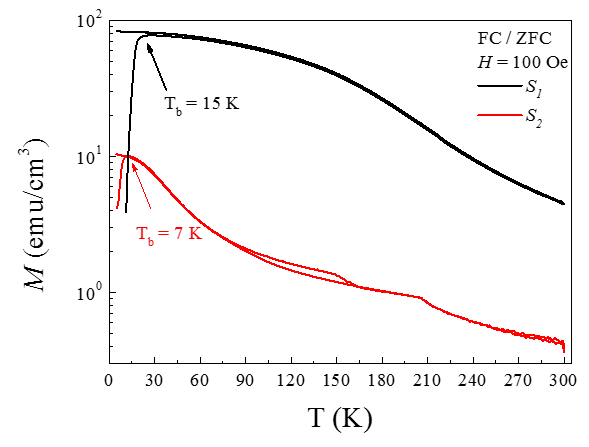}
\caption{The temperature dependencies of magnetization (FC / ZFC) of the samples $S_1$ and $S_2$ measured at the field $H=100$ Oe.}
\label{Fig7}
\end{figure}

The SQUID magnetometry was used for the investigation of the magnetic properties. The experiments were carried out at Delft University of Technology (The Netherlands) using the MPMS Quantum Design magnetometer. The magnetization loops $M(H)$ measured at room temperature are presented in Fig. \ref{Fig6}. Both curves do not reach the saturation magnetization in the in-plane field of $H= 50$ kOe and do not show any remanent magnetization and coercivity field. Thus, one can conclude that the samples are superparamagnetic. The magnetization of the sample $S_1$ is three times higher than that of the sample $S_2$ in the maximal measured field $H = 50$ kOe.

Additionally we measured the temperature dependence of the magnetization $M(T)$ in field-cooling (FC) and zero-field-cooling (ZFC) modes. The FC mode implies an application of constant magnetic field $H = 100$ Oe to the sample kept at a temperature far above a certain characteristic blocking temperature $T_b$ and cooling the sample in this field to $T < T_b$ while recording the magnetization. The ZFC mode consists of the following sequential steps: cooling the sample in zero field to $T < T_b$, applying the small field $H = 100$ Oe, and heating the sample to $T > T_b$ while measuring $M$. ZFC and FC curves overlapping down to the temperatures of 30 K for both samples $S_1$ and $S_2$ (Fig. \ref{Fig7}). The splitting between the FC and ZFC magnetization curves matches the inflection point of the ZFC curves, at which the largest number of nanoparticles is thermally unblocked. Then, the blocking temperature of the system can be found as an inflection point of ZFC magnetization curve (Fig. \ref{Fig7}). We estimated the following values of the blocking temperature: $T_{b_1} = 15$ K for the sample $S_1$ and $T_{b_2} = 7$ K for the sample $S_2$. Furthermore, broad maxima at ZFC curves reflects wide distribution over sizes of particles, as well as the interaction between them \cite{knobel2008superparamagnetism}.

\section{Discussion}
\label{S:4}

The electrical resistivity experiment shows, that the transition from a high-Ohmic to low-Ohmic state takes place in the rather narrow range of semiconductor carbon layer thicknesses from $h_c = 1.0$ nm to $h_c = 1.6$ nm (Fig. \ref{Fig1}a). To understand the origin of this transition and influence of the carbon layer thickness on the associated magnetic properties of the multilayered system, we focused on the extreme cases of study: below ($h_c = 0.4$ nm for $S_1$) and above ($h_c = 1.8$ nm for $S_2$) the threshold. The detailed investigation of the transition mechanisms of the electrical resistivity will be discussed elsewhere. According to the electrical resistance measurements (Fig. \ref{Fig1}b,c), we can conclude, that two different regimes of the conductivity of sample $S_1$ can be connected with two mechanism of the charge hopping: the Shklovskii-Efros law of tunneling between the nanoparticles at temperatures ($T>135$ K) and the Mott law of charge hopping in the amorphous carbon, where Mott law hopping takes place \cite{bucker1973preparation}. Therefore, only one regime of the Mott law conductivity can be found in the sample $S_2$, what confirms the formation of the continuous carbon layer. These results confirms, that in case of the sample $S_2$ the thickness of carbon of $h_c =1.8$ nm is enough to form the continuous carbon layer that disconnects metallic magnetic nanoparticles in neighboring layers, while in the sample $S_1$ carbon layer of thickness of $h_c =0.4$ nm is not continuous but rather formed by randomly distributed C atoms. 

According to the results of the GISAXS experiments it is possible to construct the three-dimensional structural model of the inhomogeneous magnetic multilayers {[(Co$_{40}$Fe$_{40}$B$_{20}$)$_{34}$(SiO$_2$)$_{66}$]/[C]}$_{47}$. We assume that every MI layer is a monolayer of the spheroidal nanoparticles with vertical radius $r_v$ and horizontal radius $r_h$ and volume $V=\frac{4}{3}\pi r_v r_h^2$. For a two-dimensional hexagonal monolayer of the spheroidal nanoparticles embedded into the insulator layer one can obtain:

\begin{equation}
\frac{2\pi}{3}r_v r_h^2 = \chi \frac{\sqrt{3}}{2} l^2 r_v,
\label{eq}
\end{equation}
where $l$ is the average interparticle distance in the sample plane, $\chi$ is the volume concentration of metal in the insulator matrix. Assuming that each MI layer consists of a monolayer of nanoparticles, $r_v=h_{MI}/2$ and from the Equation \ref{eq} yields:

\begin{equation}
r_h =l \sqrt{\frac{3\sqrt{3}}{4\pi}\chi}.
\label{eq2}
\end{equation}

The value $r_h$ calculated according to the Equation \ref{eq2} is presented in the Table \ref{table}. As one can see from the Table \ref{table}, the average lateral interparticle distance $l$, and in-plane size of the nanoparticles $r_h$ differs not so much for the samples $S_1$ and $S_2$ ($\chi$ = const in Equation \ref{eq2}). However, the vertical size and, consequently, volume of spheroidal nanoparticles is two times larger for $S_1$ compared to $S_2$. Thus, the nanoparticles in sample $S_1$ are elongated in vertical direction, while in sample $S_2$ the nanoparticles are flattened in the sample plane. Non-planar magnetic anisotropy of the magnetic nanogranules was previously observed in similar metal-insulator systems without interlayers FeCoZr-CaF$_2$ \cite{kasiuk2014growth}. Formation of the continuous carbon interlayer in sample $S_2$ prevents the coalescence of nanoparticles from the neighboring MI layers inducing the in-plane shape anisotropy. On the other hand, the magnetometry data also exhibit significant changes of the magnetic properties of the sample $S_1$ compared to the sample $S_2$. A blocking temperature for non-interacting superparamagnetic nanoparticles is given by formula \cite{tb}:

\begin{equation}
T_b=\frac{KV}{25k_B},
\label{tb}
\end{equation}

where $K$ is the anisotropy constant, $V$ is the nanoparticle volume and $k_B$ is Boltzmann constant. Assuming that contribution of the shape anisotropy of nanoparticles is weak compared to the magnetocrystalline anisotropy of alloy we consider $K$ the same for the both samples. Using Eq. \ref{tb} we obtain $T_{b_1}/T_{b_2} = 2.07$. The same ratio measured by SQUID is $T_{b_1}/T_{b_2} = 2.10 \pm 0.17$. Thus we can conclude that difference in blocking temperatures caused mainly by the different volume of Co$_{40}$Fe$_{40}$B$_{20}$ nanoparticles in the samples $S_1$ and $S_2$. This assumption is also supported by the fact that magnetization $M$ of the sample $S_1$ is about two times higher than magnetization of the sample $S_2$.

In the most simple model of the superparamagnetic system only particle size distribution is taken into account, while dipole-dipole interactions and magnetic anisotropy of the individual grains are ignored. Then, the magnetization $M$ as a function of field $H$ of the superparamagnetic ensemble with a particle volume distribution $f(V, \Delta V)$ is expressed by \cite{bean1959}:

\begin{equation}
M(H)=M_s \chi \int_0^\infty  L\Bigg(\frac{M_s V H}{k_B T}\Bigg) f(V, \Delta V)\, dV + \beta_{pm}H,\\
\label{mh}
\end{equation}
$$L(x) = coth(x)-1/x,$$

where $M_s$ is the saturation magnetization of ferromagnetic material, $L(x)$ is the Langevin function, and $\beta_{pm}$ is the paramagnetic term that was introduced to compensate the linear behavior of the magnetization in high fields. Assuming  $M_s =1000$ emu/cm$^3$ for Co$_{40}$Fe$_{40}$B$_{20}$ \cite{chen2012magnetic}, log-normal particle size distribution with mean volume $V$ and fitting the standard deviation $\Delta V$ one can obtain a reasonable fit of the magnetization curves for the samples $S_1$ and $S_2$ (Fig. \ref{Fig6}). The obtained deviation $\Delta V = 4.1$ nm$^3$ for the sample $S_1$ and $\Delta V = 5.0$ nm$^3$ for the sample $S_2$. We attribute the linear paramagnetic term $\beta_{pm}$ to the oxide phases of Co and Fe forming in the SiO$_2$ matrix as it was shown by XANES \cite{domashevskaya2007xanes}.

\section{Conclusion}
\label{S:5}

The presence of undamaged carbon provides the Mott law mechanism of the hopping conductivity of the sample $S_2$. On the other hand, at room temperature the conductivity of sample $S_1$ is carried out with the Shklovskii-Efros law tunneling between metallic nanoparticles, which mechanism appears to be more efficient than the Mott mechanism. The electrical resistance measurements confirmed the formation of the continuous carbon layer only in the case of sample $S_2$.

Using Grazing Incidence Small Angle X-ray Scattering, we studied the in-plane and out of plane structure of the multilayered metal-insulator/semiconductor system [(Co$_{40}$Fe$_{40}$B$_{20}$)$_{34}$(SiO$_2$)$_{66}$]/[C]$_{47}$ with two thicknesses of semiconductor layers $h_c = 0.4$ nm and $h_c = 1.8$ nm. The magnetic properties of this system measured by SQUID are explained by the morphology of metal-insulator layer which is influenced by the morphology of the carbon layer. In the investigated samples formation of the continuous amorphous carbon layer led to the reduction of the average volume of nanoparticles. Volume of magnetic grains in sample $S_2$ is two times smaller compared to the sample $S_1$, what explains why the magnetization and the blocking temperature of $S_2$ are lower then those of the sample $S_1$.	
In contrast to the similar systems [(Co$_{40}$Fe$_{40}$B$_{20}$)$_{50}$(SiO$_2$)$_{50}$]/[$\alpha$-Si]$_{60}$ with amorphous silicon interlayers we did not observe additional indirect exchange interaction between nanoparticles through the semiconductor layer.

The size and shape of nanoparticle are strongly dependent on the integrity of the carbon underlayer what is well-consisted with the previous results on [(Co$_{45}$Fe$_{45}$Zr$_{10}$)$_{35}$(Al$_2$O$_3$)$_{65}$]/[$\alpha$-Si]$_{36}$ structures \cite{Dyadkina2014}. Thus, the wetting of carbon by metal alloy leads to the reduction of the size of magnetic nanoparticles in direction perpendicular to the sample plane, and, consequently, volume and associated magnetic properties, such as magnetization and blocking temperature. The linear contribution to the magnetization $M(H)$ dependence indicates the existence of the paramagnetic oxide shells around the nanoparticles.

\subsection*{Acknowledgements}
V. Ukleev thanks Swedish Institute via Uppsala University for the financial support. We are thankful to ESRF and TU Delft for provided measurement possibilities. The work was supported by the subsidy of the Ministry of Education and Science of
the Russian Federation No. 14.616.21.0004. 





\end{document}